\documentclass[preprint2]{aastex}

\shorttitle{QPO in 1H~0707-495}
\shortauthors{Pan et al.}

\usepackage{multirow}
\usepackage{threeparttable}
\usepackage{rotating}
\usepackage{natbib}
\usepackage{subfigure}
\usepackage[usenames,dvipsnames]{xcolor}

\usepackage{multirow}
\usepackage{multicol}

\begin{document}

\title{Detection of a possible X-ray Quasi-periodic Oscillation in the Active Galactic Nucleus 1H~0707-495}

\author{Hai-Wu Pan\altaffilmark{1,2}, Weimin Yuan\altaffilmark{1,2}, Su Yao\altaffilmark{1,2}, Xin-Lin Zhou\altaffilmark{1}, Bifang Liu\altaffilmark{1,2}, Hongyan Zhou\altaffilmark{3,4} \& Shuang-Nan Zhang\altaffilmark{5,1}}

\altaffiltext{1}{Key Laboratory of Space Astronomy and Technology, National Astronomical Observatories, Chinese Academy of Sciences, 20A Datun Road, Chaoyang District, Beijing, China; panhaiwu@nao.cas.cn, wmy@nao.cas.cn}
\altaffiltext{2}{University of Chinese Academy of Sciences, School of Astronomy and Space Science, Beijing 100049, China}
\altaffiltext{3}{Polar Research Institute of China, 451 Jinqiao Road, Pudong, Shanghai 200136, China}
\altaffiltext{4}{Key Laboratory for Research in Galaxies and Cosmology, The University of Sciences and Technology of China, Chinese Academy of Sciences, Hefei, Anhui 230026, China}
\altaffiltext{5}{Key Laboratory of Particle Astrophysics, Institute of High Energy Physics, Chinese Academy of Sciences, Beijing 100049, China}

\begin{abstract}
Quasi-periodic oscillation (QPO) detected in the X-ray radiation of black hole X-ray binaries (BHXBs) is thought to originate from dynamical processes in the close vicinity of the black holes (BHs), and thus carries important physical information therein. Such a feature is extremely rare in active galactic nuclei (AGNs) with supermassive BHs.
Here we report on the detection of a possible X-ray QPO signal with a period of 3800\,s at a confidence level $>99.99\%$ in the narrow-line Seyfert 1 galaxy (NLS1) 1H~0707-495 in one data set in 0.2-10\,keV taken with {\it XMM-Newton}. The statistical significance is higher than that of most previously reported QPOs in AGNs. The QPO is highly coherent (quality factor $Q=\nu/\Delta\nu \geqslant 15$) with a high rms fractional variability ($\sim15\%$).
A comprehensive analysis of the optical spectra of this AGN is also performed, yielding a central BH mass $5.2\times10^6\,M_{\odot}$ from the broad emission lines based on the scaling relation. The QPO follows closely the known frequency-BH mass relation, which spans from stellar-mass to supermassive BHs.
The absence of the QPO in other observations of the object suggests it a transient phenomenon.
We suggest that the (high-frequency) QPOs tend to occur in highly accreting BH systems, from BHXBs to supermassive BHs. Future precise estimation of the BH mass may be used to infer the BH spin from the QPO frequency.
\end{abstract}


\section{INTRODUCTION}

Active galactic nuclei (AGNs), powered by black hole (BH) accretion with BHs of $10^5-10^9\,M_{\odot}$ at the center of galaxies, are thought to be scaled-up versions of black hole X-ray binaries (BHXBs) \citep{M10,G12,M15,S15}. A compelling line of evidence for this postulation is the striking similarity in the variability of the X-ray radiation between AGNs and BHXBs (e.g., \citealp{U02,M03,M06,M10,V03,V11}).
One characteristic and enigmatic feature of the variable X-rays is Quasi-Periodic Oscillation (QPO), which has been observed in the X-ray light curves of dozens of BHXBs (e.g., \citealp{S01,R02}). X-ray QPOs were found to have two types, low-frequency QPOs (LFQPOs) and high-frequency QPOs (HFQPOs) (\citealp{R06}, hereafter RM06). HFQPOs sometimes occur in pairs with a constant frequency ratio of 3:2, and unlike LFQPOs, their frequencies ($f_{\rm QPO}$) are stable despite significant variations in fluxes \citep{R02}. This indicates that HFQPO is likely a fundamental property that might be linked to the mass (and spin) of the BH. Based on only three BHXBs with measured BH masses, RM06 suggested an inverse linear relation between $f_{\rm QPO}$ and BH mass, which was later extended to intermediate-mass BH (IMBH) in ultraluminous X-ray sources (ULXs) by \citet{A2004}.
A number of HFQPO models have been proposed in the literature (see \citealp{L2009} and \citealp{B14} for reviews and references therein).

In AGNs with supermassive BHs (SMBHs), however, QPOs are rarely detected. So far, there is only one widely accepted case in which a significant QPO was unambiguously detected in the Narrow-Line Seyfert 1 (NLS1) galaxy RE~J1034+396 with $f_{\rm QPO}\simeq2.7\times10^{-4}$\,Hz \citep{G08,A14,H14}. Recently, A $\sim 200$\,s X-ray QPO was detected in the transient X-ray source Swift~J164449.3+573451, thought to be of tidal disruption of a star by a dormant SMBH \citep{R12}. By extending BH masses to the SMBH range, \citet{Z15} suggested that the $f_{\rm QPO}-M_{\rm BH}$ scaling relation is universal spanning $\sim6$ orders of magnitude from stellar-mass BHs to SMBHs.

Besides RE~J1034+396, possible detections of X-ray QPOs were also reported in a few other AGNs, however, at relatively low levels of significance. A $\sim 3.8$\,hr QPO in 2XMM~J123103.2+110648 was reported and suggested to be a LFQPO \citep{L13}, and a $\sim 2$\,hr HFQPO was also reported in MS~2254.9-3712 \citep{A15}.

In this letter, we report on the discovery of a significant QPO signal in one {\it XMM-Newton} observation of 1H~0707-495, a nearby (redshift 0.04), typical NLS1 \citep{VV10}. We also re-examine the BH mass of 1H~0707-495 by analysing its available optical spectroscopic data, and compare the QPO with the $f_{\rm QPO}-M_{\rm BH}$ relation.
A cosmology with $H_{\rm 0}=70 \rm km s^{-1} Mpc^{-1}$, $\Omega_{\rm m}=0.3$ and $\Omega_{\rm \Lambda}=0.7$ is adopted.

\section{X-RAY QUASI-PERIODIC OSCILLATION}

1H~0707-495 was observed with {\it XMM-Newton} with an exposure of $\sim 100$\,ks on February 6 2008 in the full frame imaging mode (Obs ID: 0511580401).
We follow the standard procedure to reduce the data and extract science products from the observation data files (ODFs) using the {\it XMM-Newton} Science Analysis System (SAS, version 13.0.0). Only good events (single and double pixel events, i.e. PATTERN $\leq $ 4 for PN or $\leq $ 12 for MOS) are used. Source events are extracted from a 40-arcsec circular region, and background events from a source-free circle with the same radius on the same CCD chip. X-ray light curves are constructed for all the three EPIC detectors in the 0.2-10\,keV energy band with a binsize of 100\,s, and are corrected for instrumental factors using EPICLCCORR.

\begin{figure}
\epsscale{.85}
\plotone{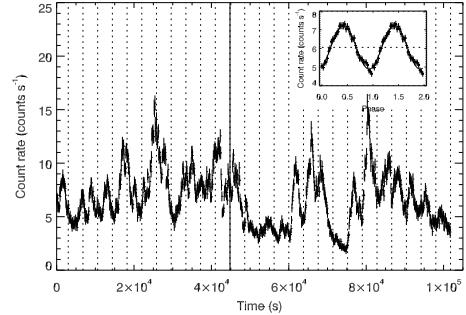}
\caption{{\it XMM-Newton} light curve of 1H~0707-495, which is extracted and combined from the PN, MOS1 and MOS2 detectors in 0.2-10\,keV with a binsize of 100\,s. The light curve is divided into two segments by the solid line, and only the second segment is used in this letter. The dotted vertical lines show the expected periodicity of 3800\,s, which can be seen from the light curve even by eye. {\it Upper Right Inset}: the folded light curve with a period 3800\,s. Errors are propagated from the unfolded curve. The best-fit sinusoid is showed as the solid line and the mean count rate as the dotted line. Two cycles are plotted for clarity.\label{fig1}}
\end{figure}

The combined PN+MOS1+MOS2 light curve is shown in Figure \ref{fig1}.
Even by eye, a periodicity can be clearly seen. We search for periodicity by calculating the root-mean-square (rms) amplitudes of the light curve folded with various periods as a function of period. A strong peak appears at around 3800\,s. It is also found that this peak is the strongest if the latter half of the light curve is used only, with a quasi-period of $3800\pm170$\,s (full-width at half-maximum).
We thus take an operational cut as indicated in Figure \ref{fig1} (solid vertical line) and consider the last $\sim 55$\,ks segment only hereafter, in order to achieve the highest confidence level for the periodic signal. This segment has a mean count rate of 6.1\,counts\,s$^{-1}$ and fractional rms variability 41\%, while the whole light curve gives 6.7\,counts\,s$^{-1}$ and 37\%, respectively.
The folded light curve with a period 3800\,s is also showed in Figure \ref{fig1} (inset), with the best-fit sinusoid model.

\begin{figure}
\epsscale{.85}
\plotone{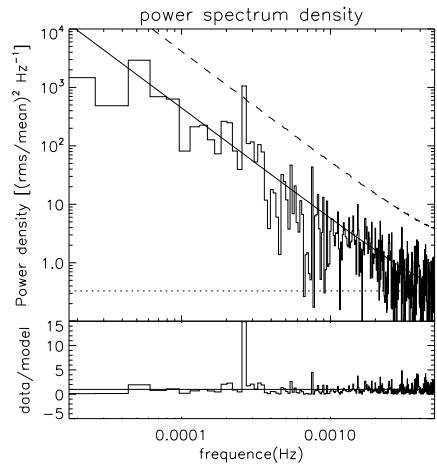}
\caption{Power spectrum density of the X-ray light curve of 1H~0707-495. The solid line represents the best-fit power-law, and the dashed line represents the 99.99\% confidence level. The dotted horizontal line shows the expected level of the Poissonian noise. The data/model residuals are shown in the lower panel. A statistically significant peak is clearly present at $\sim2.6\times10^{-4}$\,Hz. \label{fig2}}
\end{figure}

The power spectrum density (PSD) of the light curve is computed as the modulus-squared of the discrete Fourier transform (DFT), and the (rms/mean)$^2$ normalisation is chosen \citep{V03b}. The PSD for the last 55\,ks segment is plotted in Figure \ref{fig2}, and a strong peak at $2.6\times10^{-4}$\,Hz is evident. The continuum (red noise) is well fitted with a power-law of a slope $-1.90$ ($\chi^2=301/282~dof$, in the log-log space). This signal is even more prominent in the residuals plot (data/model). To test its statistical significance, a Monte Carlo technique is applied following \citet{R12} and \citet{L13}. We simulate one million light curves using the method of \citet{T95} from a PSD, which is assumed to be the above best-fit power-law. Thus the distribution of the variability power at each fourier frequency can be obtained from the PSDs of the simulated light curves. The dashed line in Figure \ref{fig2} represents the 99.99\% significance level, indicating a significant QPO. Using the data of PN camera and the combined MOS cameras individually leads to the same result with the QPO significance levels $>99.99\%$ in both cases.

To test the QPO in RE~J1034+396, \citet{G08} used a method developed by \citet{V05b}. This method was later improved significantly by \citet{V10} to be an even more stringent test based on Bayesian statistics, in combination with PSD fitting. We also adopt this improved method here. Instead of a simple power-law ($H_0$ model), a bending power-law ($H_1$ model) is used to fit the PSD continuum. The low-frequency slope is fixed at -1, while the high-frequency slope is fitted to be $-2.4\pm0.22$, with a bending frequency of $1.4\times10^{-4}$\,Hz. Then Markov Chain Monte Carlo simulations are performed, resulting a small posterior predictive $p$-value of $2.5(\pm0.5)\times10^{-3}$ for the significant outlier at the frequency $\sim2.6\times10^{-4}$\,Hz [$p=3.9(\pm0.62)\times10^{-3}$ for the PN light curve and $p=4.0(\pm0.63)\times10^{-3}$ for the combined MOS light curve, respectively]. This strongly indicates the presence of the QPO.

This periodicity of $\sim2.6\times10^{-4}$\,Hz is highly coherent and confined to only one frequency bin in our analysis. Considering the frequency resolution for the observation ($1/T=1.75\times10^{-5}$\,Hz), the quality factor ($Q=\nu/\Delta\nu \geqslant 15$) is high. The rms fractional variability in the QPO is $\sim15\%$. Compared to the rare previously reported QPOs in AGNs in the literature, the QPO in 1H~0707-495 has a significance level at least similar to or even higher than most of the others but only somewhat lower than that in RE~J1034+396 \footnote{The $p$-value for RE~J1034+396 is $\sim5\times10^{-4}$, if only the second segment of its light curve is used, following \citet{G08}.}.
The significance of this QPO would be reduced to $<90\%$ if the whole observation duration is considered.
We also analyzed data of other 14 {\it XMM-Newton} observations of 1H~0707-495 with exposure time longer than 40\,ks, but no significant QPO is found. Thus the QPO appears to be a transient feature; similar to that found in RE~J1034+396 \citep{G08,M11}

\section{BLACK HOLE MASS ESTIMATION AND THE $f_{\rm QPO}-M_{\rm BH}$ RELATION}

\begin{figure}[!ht]
\centering
    \includegraphics[width=0.8\columnwidth]{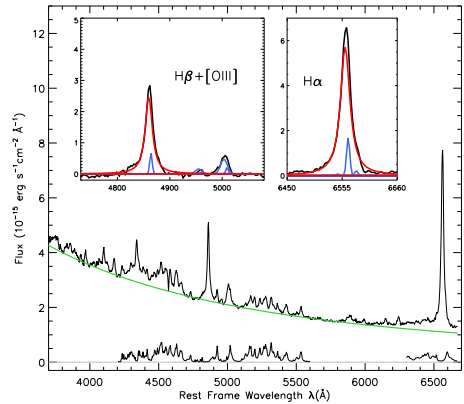}
    \caption{CTIO spectrum of 1H\,0707$-$495. The continuum modeled with a power-law is plotted in green. 
    The best-fit Fe{\scshape~ii} emission is represented by the lower black curve.
    The residual emission line spectra (black line) after subtracting the power-law continuum and the Fe{\scshape~ii} model is shown in the insets for the H$\beta+$[O\,{\scshape iii}] and H$\alpha$ regions, respectively (red: broad lines; blue: narrow lines). See text for details.
    \label{fig3}
    }
\end{figure}

Using the empirical virial method based on optical spectroscopic data, the BH mass of 1H~0707-495 has been estimated in previous studies, as $2_{-1}^{+4}\times10^6\,M_{\odot}$ from a 6dF spectrum \citep{B03} and $4\times10^6\,M_{\odot}$ \citep{D15} from CTIO spectrum taken by \citet{L04}.
We re-analyze both the spectra, following the procedure as described in \citet{Y15} to fit the optical spetra.
Only the CTIO spectrum is shown in Figure \ref{fig3} for demonstration.
The flux of the 6dF spectrum is calibrated using the CTIO spectrum.
The continuum is modeled with a power law and the Fe{\scshape~ii} emission is modeled using the \citet{V04} templates.
The Balmer lines are deblended into a broad and a narrow component,
and the former is modeled with a Lorentzian profile.
The narrow component is modeled by a Gaussian with FWHM fixed at the CTIO spectral resolution ($\sim$330\,km\,s$^{-1}$, \citealp{L04}).
The flux ratio of the [O\,{\scshape iii}]\,$\lambda\lambda$4959,\,5007 doublet is fixed at the theoretical value of 1:3 while each of them is fitted with two Gaussians, one for the line core with FWHM fixed to the narrow Balmar line widths, the other for the blueshifted wing.
The widths of the broad H$\beta$ line are fitted to be 1002\,km\,s$^{-1}$ (CTIO) and 1054\,km\,s$^{-1}$ (6dF), which agree with each other within mutual errors.
The narrowness of the broad H$\beta$ line, as well as the strong Fe{\scshape~ii} emission are characteristic of NLS1 AGN.
The BH mass is estimated using the broad H$\beta$ line width and the monochromatic luminosity at 5100\,\AA~($\lambda L_{5100}=4.0\times10^{43}$\,erg\,s$^{-1}$) \citep{V06}, giving $M_{\rm BH}=5.2\times10^6\,M_{\odot}$ (CTIO) and $5.7\times10^6\,M_{\odot}$ (6dF), respectively.

Given the large uncertainty $\sim0.5$\,dex of this method \citep{V06}, the estimated BH mass is consistent with previous results. We consider our spectral analysis to be more comprehensive and rigorous compared to those in previous work, and thus $M_{\rm BH}=5.2(\pm0.5\,dex)\times10^{6}\,M_{\odot}$ to be the best-estimate of the BH mass in 1H\,0707$-$495.
Adopting a bolometric luminosity of $3.6\times10^{44}$\,erg\,s$^{-1}$  estimated from $9\times \lambda L_{5100}$ \citep{K00}, the Eddington ratio is 0.5, which is typical of NLS1.

\begin{figure}[!ht]
\epsscale{0.90}
\plotone{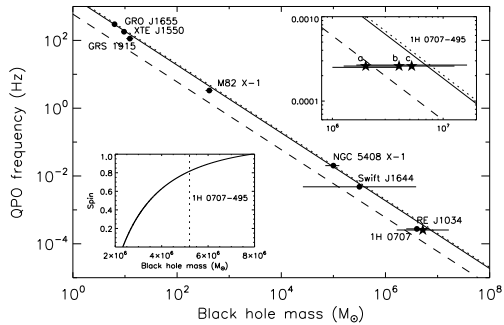}
\caption{Relation between QPO frequency and BH mass. This is an updated version of Figure 1 from \citet{Z15}, where objects with significant QPO detections are plotted (dots), together with the newly detected QPO in 1H~0707-495 (star). The solid line represents the extrapolation of the relation derived in RM06, based on the three BHXBs. The dotted line and dashed line represent the relations derived from the model of 3:2 resonance \citep{K02} with the spin parameter a=0.998 (dotted) and a=0 (dashed), respectively. {\it Upper right inset}: A zoom-in comparison of 1H~0707-495 with the model predictions, with various available BH mass estimates. a and b denote the mass from \citet{B03} and \citet{D15} respectively, while c denotes our best-estimated mass derived in this paper. {\it Lower left inset}: the inferred BH spin for a range of BH mass values for 1H~0707-495 from the 3:2 resonance model, assuming that the QPO represents the $2\times f_{0}$ peak. \label{fig4}}.
\end{figure}

By extending the  $f_{\rm QPO}-M_{\rm BH}$ relation from stellar-/intermediate-mass BH to the SMBH regime, \citet{Z15} suggested it to be a universal relationship for astrophysical BHs at all mass scales. In Figure \ref{fig4}, we reproduce the $f_{\rm QPO}-M_{\rm BH}$ relation derived by RM06 based on three BHXBs (solid line) and overplot 1H~0707-495, as well as those BH systems with known QPOs (see Table 1 in \citealp{Z15}).
Clearly, the QPO in 1H~0707-495 conforms with the $f_{\rm QPO}-M_{\rm BH}$ relation within the BH mass uncertainty range.

\section{DISCUSSION}

\subsection{Reliability of the QPO and Comparison with RE~J1034+396}

A significant QPO signal is detected at a $>99.99\%$ confidence level in one of the X-ray light curves of the NLS1 AGN 1H~0707-495 taken with {\it XMM-Newton}. The QPO frequency ($\sim2.6\times10^{-4}$\,Hz) and the estimated mass of the central BH ($M_{\rm BH}=5.2\times10^{6}\,M_{\odot}$) conform with the previous $f_{\rm QPO}-M_{\rm BH}$ relation.
We note that the same data set was also analyzed by \citet{G12} in a study of the X-ray timing properties of a large sample of AGN, but no significant QPO was reported.
This might be ascribed to that the data of the whole observation duration was used in their work, which led to reduced significance of the signal ($<90\%$, most likely due to a shift of the QPO phase). The same situation also occurred in the case of RE~J1034+396, in which the QPO was significant in only part of the light curve, and the significance became much lower when the whole light curve was used \citep{G08,M11}.
For 1H~0707-495 there are 15 {\it XMM-Newton} observations with good quality light curves, totaling 1200\,ks exposure time. This may be equivalent to that the QPO is detected in only one out of effectively $\sim 1200$\,ks$/55$\,ks$\backsimeq22$ searches.
Considering all the non-detections would lower the overall significance of the detection in the statistical sense, giving only a null probability $p\sim22\times (2.5\times10^{-3})\sim 0.05$.
However, this is the most conservative case and should only be considered as a limit, since it is known that a QPO does not always repeat in every observation.
QPOs in AGNs may well be a transient phenomenon (e.g., the data suggest a duty cycle $\sim5\%$ in the case of 1H~0707-495).
Furthermore, the fact that the QPO in 1H~0707-495 follows closely the $f_{\rm QPO}-M_{\rm BH}$ relation supports its genuineness.

The QPO frequency in 1H~0707-495 is very close to that in RE~J1034+396 ($\sim2.7\times10^{-4}$\,Hz), which was argued to be a HFQPO \citep{Z10,Z15}.
This is not surprising given their similar BH masses ($M_{\rm BH}=4_{-2}^{+3}\times10^6\,M_{\odot}$ for RE~J1034+396 and $5.2\times10^{6}\,M_{\odot}$ for 1H~0707-495).
Moreover, the two AGNs are similar in several ways, e.g., the NLS1 classification \citep{VV10}, the high Eddington ratios ($L_{\rm Bol}/L_{\rm Edd}=1.25$ for RE~J1034+396 and 0.5 for 1H~0707-495), the X-ray spectra with a strong soft excess below 1\,keV, as well as the PSD shapes \citep{G12}.
This suggests that the HFQPOs tend to occur in this type of AGNs.

\begin{figure}[!ht]
\centering
\includegraphics[width=0.9\columnwidth]{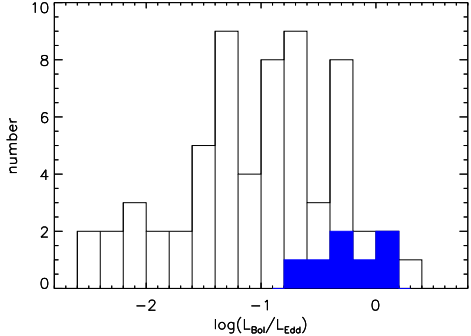}
\caption{Distribution of the Eddington ratios of the BH accretion systems with HFQPO detected (blue), including BHXBs, ULX (M82~X-1), TDE (Swift~J164449.3+573451) and AGNs (1H~0707-495 and RE~J1034+396). As a comparison, also plotted is the distribution of the Eddington ratios of AGNs from the \citet{G12} sample in which QPO was searched for but not detected. The Kolmogorov-Smirnov test gives a probability $p=0.2\%$ that the two distributions are drawn from the same population. \label{fig5}}.
\end{figure}

\subsection{Do HFQPOs Tend to Occur at the Highest Accretion State?}

HFQPOs in BHXBs are observed only when the systems are at the very high state with high accretion rates \citep{R06,L2009}.
Interestingly, the two AGNs having significant QPO detections, RE~J1034+396 and 1H~0707-495, are both NLS1s, which are thought to be an AGN analogue of BHXB at the very high state.
Figure 5 shows the distribution of the Eddington ratios for all the BH accretion systems with reliable QPO detections (as shown in Figure \ref{fig4}) from BHXBs to AGNs. For the three BHXBs, the Eddington ratios are calculated from the average bolometric luminosities at which the QPO occurred from \citet{R02} and \citet{B06}.
For the tidal disruption event (TDE) Swift~J164449.3+573451 in which a QPO was found, the Eddington ratio is believed to be around unity or even higher since its observed luminosity is highly super-Eddington \citep{R12}.
The Eddington ratio of M82~X-1 was suggested to be 0.8 \citep{P14}.
As a comparison, we overplot the Eddington ratio distribution for the AGN sample of \citet{G12}, in which QPOs were searched for but not found.
Clearly the two distributions differ significantly; a two-sided Kolmogorov-Smirnov test yields a small $p$-value of 0.2\%. This result suggests that the HFQPOs tend to occur at the highest accretion state of BH accretion systems.

\subsection{Origins of HFQPO and Possible Link with BH Spin}

The origin of the HFQPO is unclear, though a number of models have been proposed in the literature, e.g., the relativistic precession model (RPM, \citealp{S99}), resonance model (RM, \citealp{A01}), acoustic oscillation modes in pressure-supported accretion tori (AOM, \citealp{R03}), instability at disk-magnetosphere interface (IDI, \citealp{LN04}), accretion-ejection instability in magnetized disks (AEI, \citealp{T99}), and global corotational instability of non-axisymmetric g-mode or p-mode trapped in the innermost region of the accretion disk (g-/p-mode models; \citealp{L03}). All these models can explain the 3:2 frequency ratio of the twin-peak QPOs. The $f_{\rm QPO}-M_{\rm BH}$ relation is also predicted in all except the g-mode model \citep{S08}. Only the AEI and p-mode models can explain why the HFQPOs occur exclusively at the very high accretion state. A common issue for the RPM, RM and AOM models is a lack of underlying physical mechanisms at work, e.g., how orbiting hot spots survive in a differentially rotating disk, how the resonances are produced, etc (see \citealp{L2009,B14} for brief reviews and references therein). Discussion concerning these models is beyond the scope of this paper. Nevertheless, here we briefly address the explanation and implication of our results in the framework of the resonance model.

In the resonance model, the HFQPOs are suggested to be associated with the frequencies of three fundamental oscillation modes of disk fluid elements around a BH: the Kepler orbital motion, the radial and vertical (to the orbital plane) epicyclic oscillation modes \citep{A01}. These frequencies naturally scale inversely and linearly with the BH mass, with a dependence on the BH spin \citep{N98}.
Assuming that the observed 3:2 twin-peak QPO frequencies correspond to the vertical and radial epicyclic frequencies as in \citet{K02}, the predicted $f_{\rm QPO}-M_{\rm BH}$ relation is well consistent with the fitted relation from RM06 (solid line in Figure \ref{fig4}), which corresponds to a high BH spin close to the maximum value \citep{Z15}. To illustrate the effect of BH spin, the $f_{\rm QPO}-M_{\rm BH}$ relation assuming two extreme spin values a=0 (no spin; dashed) and a=0.998 (maximum spin; dotted) are overplotted in Figure \ref{fig4} for the $2\times f_{0}$ frequency (the lower frequency of the 3:2 twin-peak QPO). A zoom-in comparison with the results of 1H~0707-495 is shown in the upper-right inset of Figure \ref{fig4}, where the BH mass estimates in this and previous work are over-plotted.
As can be seen, the model can well reproduce the observed QPO frequency for a wide range of BH spin, given the relatively large uncertainty range of $M_{\rm BH}$. The inferred spin values from a range of BH masses are shown in the lower-left inset in Figure \ref{fig4}. As one immediate inference, the allowed BH mass in 1H~0707-495 is in the range $(2-8)\times10^6\,M_{\odot}$ (corresponding to from a=0 to a=0.998). Our best-estimate $M_{\rm BH}=5.2\times10^{6}\,M_{\odot}$ indicates a moderately high BH spin $\sim0.8$.
It should be noted that \citet{F09} suggested a high BH spin for 1H~0707-495 based on its relativistic Fe line and reflection spectra, while \citet{D15} argued for a low spin.
In general, NLS1 AGN are suggested to have averagely low or moderate BH spins as a population, as found by \citet{Liu15} from the relativistic Fe line profile of the stacked spectra.
Clearly, a precise BH mass measurement, e.g., via the reverberation mapping method with which an accuracy $\sim30\%$ could be reached \citep{P04}, is needed for constraining the BH spin in such AGNs with detected QPOs.

\acknowledgments
W.Y. thanks S. Vaughan, P. Uttley and C. Jin for helpful discussion. S. Vaughan is also thanked for making the Bayesian test code available, and C. Jin and K. Leighly for providing the CTIO spectral data. This work is supported by the National Natural Science Foundation of China (Grant No.11473035), and the Strategic Priority Research Program ``The Emergence of Cosmological Structures'' of the Chinese Academy of Sciences (Grant No. XDB09000000). This work is based on observations obtained with {\it XMM-Newton}, an ESA science mission with instruments and contributions directly funded by ESA Member States and NASA. This research has made use of the NASA/IPAC Extragalactic Database (NED).

\end{document}